\documentstyle[eqsecnum,aps,prd]{revtex}
\begin{document}
%---substitution
%-------a):basics
%---------a1):environments
\newcommand{\beq}{\begin{equation}}
\newcommand{\eeq}{\end{equation}}
\newcommand{\beqn}{\begin{eqnarray}}
\newcommand{\eeqn}{\end{eqnarray}}
\newcommand{\nnb}{\nonumber}
%---------a2):mathematic symbol
\newcommand{\ppp}{\partial}
\newcommand{\V}{\nabla}
\newcommand{\EEE}{\epsilon}
\newcommand{\Lie}{\mbox{$\pounds$}}    % makes the followings italic without '\mbox'
\newcommand{\ZZ}{\mbox{${\sf Z}{\hskip -0.3em}{\sf Z}$}}
\newcommand{\tri}{\triangle}
\newcommand{\triB}{\mbox{$\bm\tri$\hskip-1.8ex\raisebox{-0.7ex}{$\bar{}$}\hskip1.7ex}{}}
\newcommand{\h}{\hat}
\newcommand{\tl}{\tilde}
\newcommand{\arcsinh}{\mbox{\rm arcsinh}}
\newcommand{\arccosh}{\mbox{\rm arccosh}}
\newcommand{\base}[2]{ {\scriptstyle (} #1_{\! \ssr{#2}} {\scriptstyle )} }
\newcommand{\ssr}[1]{\scriptscriptstyle{\rm \, #1}}
\newcommand{\up}[1]{{}^{^{#1}}{\hskip-3.0pt}}
\newcommand{\w}{\omega}
\newcommand{\k}{\mbox{$\sf{K}$}}
\newcommand{\mmbox}[3]{\mbox{$\hspace{#1}\mbox{#2}\hspace{#3}$}} %
%======================================%
%<<<<<<<<<<<< TITLE PAGE >>>>>>>>>>>>>>%
%======================================%
\thispagestyle{empty}
{\baselineskip0pt
\leftline{\large\baselineskip16pt\sl\vbox to0pt{\hbox{DAMTP} 
               \hbox{University of Cambridge}\vss}}
\rightline{\large\baselineskip16pt\rm\vbox to20pt{
               \hbox{DAMTP-1999-60}
               \hbox{UTAP-325}
               \hbox{RESCEU-12/99}
               \hbox{\today}
\vss}}%
}
\vskip15mm

\begin{center}
{\large\bf Space-times which are asymptotic to certain
Friedman-Robertson-Walker space-times at timelike infinity}
\end{center}

\begin{center}
{\large Tetsuya Shiromizu${}^{a, c, d}$
\footnote{JSPS Postdoctal Fellowship for Research Abroad} and  
Uchida Gen${}^{b,c}$}
\vskip 3mm
\sl{${}^{a}$DAMTP, University of Cambridge \\ 
Silver Street, Cambridge CB3 9EW, UK \\
\vskip 5mm
${}^{b}$Department of Earth and Space Science, Graduate School of Science,\\
Osaka University, Toyonaka 560-0043, Japan \\
\vskip 5mm
${}^{c}$Department of Physics, The University of Tokyo, Tokyo 113-0033, 
Japan \\
and \\
${}^{d}$Research Centre for the Early Universe(RESCEU), \\ 
The University of Tokyo, Tokyo 113-0033, Japan}
\end{center}
\vskip 5mm
%\begin{center}
%{\it Submitted to CQG}
%\end{center}
%======================================%
%<<<<<<<<<<<<< ABSTRACT >>>>>>>>>>>>>>>%
%======================================%
\begin{abstract}
We define space-times which are asymptotic to radiation dominant
Friedman-Robertson-Walker space-times at timelike infinity and study the 
asymptotic structure. We discuss the local asymptotic symmetry  and 
give a definition 
of the total energy from the electric part of the Weyl tensor. 
\end{abstract}
%\pacno{04.20.Cv}

\section{Introduction}

According to the present observations we are living in an expanding 
universe and we know that the the space-time is approximately described by 
Friedman-Robertson-Walker(FRW) space-times. So far only the
structure of asymptotically Minkowskian space-time is 
studied. The formulation is adequate to understand the structure and evolution of 
compact objects and the stability of space-time. However, if we wish
to study the formation of very early objects such as primordial black
holes\cite{PBH}, we should define asymptotically FRW space-times. 
Thus, it is important to define 
and investigate the structure and stability of asymptotically FRW 
space-times. 

There is another reason to investigate asymptotic structure of
asymptotically FRW space-times.  
Standard inflation predicts flat FRW universe\cite{Inflation}. 
Since observations of baryon suggest the small density parameter 
$\Omega_0 \sim 0.3$, a positive cosmological constant is needed for 
the standard inflation scenario. However, the 
existence of the cosmological constant is not clear in 
aspects of observations and fundamental particle physics at present. 
A few years ago the creation of an open FRW universe aided by one bubble 
inflation has been proposed\cite{Bubble}. Unfortunately, the 
scenario demands a fine-tuning on the potential of inflaton. 
Recently, Hawking and Turok proposed a scenario that an open FRW universe can be 
created from `nothing'\cite{Open}. Since the Hawking-Turok 
instanton has timelike singularity, the issue is 
still under debate\cite{Debate}. 
However, their scenario of the quantum creation of an open universe is 
very attractive because the model needs no strong fine-tuning. 
Thus, it is important to obtain the basic structure of 
asymptotically open FRW universe in order to investigate the nature 
of the instanton.  

In this paper, we will define space-times which are asymptotic to 
radiation dominant FRW space-times at timelike infinity
(for brevity, we will call them by AFRWTI space-times hereafter), 
except for closed FRW universe. This is 
because closed FRW universes terminate at the big-crunch and the 
notion of AFRWTI space-times is lost at the singularity.   
We use the mathematical tool which 
was developed by Ashtekar and Romano\cite{AR} for spatial 
infinity and extended by Gen and Shiromizu\cite{Gen} to timelike
infinity in the asymptotically flat space-times. 

The rest of this paper is organised as follows. In Sec. II, we will 
treat FRW space-times to look for a 
better definition of AFRWTI space-times. 
In Sec. III we define AFRWTI space-times and 
show that spacelike hypersurfaces of space-times approach conformally 
to three-dimensional Euclid space or hyperbolic space.  
In Sec. IV, we consider the asymptotic symmetry. In Sec. V, we 
define the total energy of AFRWTI space-times from the electric part of 
the Weyl tensor. Finally we give 
a summary and discuss about the relation between the total energy 
and the gravitational instanton in Sec. VI. 
In the appendix, we will give pedagogical
examples for AFRWTI space-times. We basically follow the notation of Wald's 
text\cite{Wald}. 

\section{FRW space-times}

The covariant analysis of asymptotic structure has 
been studied so far by using the conformal completion of 
space-times\cite{AH}. 
The conformal completion is useful for the investigation of 
the global causal structure and can treat simultaneously 
null and spatial infinities. However, the method leads to 
{\it a point} spatial infinity and this results in complicated 
differential structure at spatial infinity. To overcome this 
difficulty,   
Ashtekar \& Romano have proposed a new completion of space-time at 
spatial infinity and have resolved the problem\cite{AR}. They 
abandoned to investigate the causal structure and did not use 
conformal completion. Using this new 
method, we discussed a hierarchy 
structure of asymptotic gravitational fields\cite{Gen}\cite{Perng} 
like {\it peeling behaviour}\cite{Penrose} at timelike infinity and 
spatial infinity. 
We adopt the new completion to investigate asymptotic 
structure of universe in time.

Let us consider FRW space-times in order to obtain 
the essence of the new completion. The metric of spatially open FRW
space-times is given by 
%========<Equation>========%
%
\begin{eqnarray}
ds^2=a^2(\eta)[-d\eta^2 +\gamma_{ij}dx^idx^j],
\end{eqnarray}
%
%==========================%  
where $\gamma_{ab}$ is the metric of the three-dimensional unit 
hyperboloid space. The scale factor $a(\eta)$ can be written as 
$a(\eta)={\rm sinh}(\eta/\eta_0)$ and $a(\eta)={\rm
cosh}(\eta/\eta_0)-1$ for radiation and matter dominant universe, 
respectively.

We introduce a coordinate $\Omega = a^{-1}$ to define 
the asymptotic region at the timelike infinity. In terms of $\Omega$ 
the metric becomes
%========<Equation>========%
%
\begin{eqnarray}
{\hat g}_{\sf{\scriptscriptstyle{FRW}}}=-F_0^{-1}\Omega^{-4} 
\Bigl[ 1-\Omega^2+O(\Omega^3) \Bigr]
d\Omega^2
+\Omega^{-2}\gamma_{ij} dx^i dx^j,
\end{eqnarray}
%
%==========================%  
and 
%========<Equation>========%
%
\begin{eqnarray}
{\hat g}_{\sf{\scriptscriptstyle{FRW}}}=-F_0^{-1}\Omega^{-4} 
\Bigl[ 1-2\Omega+O(\Omega^2) \Bigr]
d\Omega^2
+\Omega^{-2}\gamma_{ij} dx^i dx^j,
\end{eqnarray}
%
%==========================%  
for radiation and matter dominant open FRW universe, respectively. 
In the above $F_0^{-1} =\eta_0^2$. In the coordinate system 
$(\Omega, x^i)$ the metric is singular at the timelike infinity 
$\Omega \to 0$. However, 
%========<Equation>========%
%
\begin{eqnarray}
n^a=\Omega^{-4}{\hat g}^{ab}{\hat \nabla}_b \Omega 
=-F(\partial_\Omega)^a   
\end{eqnarray}
%
%==========================%  
and 
%========<Equation>========%
%
\begin{eqnarray}
q_{ab}=\Omega^2 a^2\gamma_{ij}dx^i dx^j=\gamma_{ij}dx^idx^j.
\end{eqnarray}
%
%==========================%  
are smooth tensors at the infinity and contain all informations  
of gravitational field. Hence the gravitational field of asymptotic 
region can be investigated in terms of the pair $(n^a,q_{ab})$. 

As we can see, the leading behaviour of the metric does not depend on
the equation of state. Since the curvature dominant open universe 
approaches to Milne universe which is a submanifold 
of Minkowski space-time covered by the hyperbolic coordinate, 
we can apply directly our previous study on the timelike infinity of 
asymptotically flat space-times\cite{Gen} for spatially open FRW 
space-times. 

Next, we consider the spatially flat FRW space-times whose 
metric is given by 
%========<Equation>========%
%
\begin{eqnarray}
ds^2=a^2[-d\eta^2 +\gamma_{ij}dx^idx^j],
\end{eqnarray}
%
%==========================%  
where $ad\eta = dt$, $a=a_0(t/t_0)^{2/3\gamma}$ and 
$\gamma_{ab}$ is the metric of the 3-Euclid space. 
$\gamma =4/3, 0$ correspond to radiation and  
matter dominant universe, respectively.  

In the same way as the open FRW cases, we introduce a 
coordinate $\Omega = (a/a_0)^{-1}$ to define 
the asymptotic region at the timelike infinity.  In terms of $\Omega$ 
the metric becomes
%========<Equation>========%
%
\begin{eqnarray}
{\hat g}_{\sf{\scriptscriptstyle{FRW}}}=-F^{-1}\Omega^{-3\omega-5}d\Omega^2
+\Omega^{-2}\gamma_{ij} dx^i dx^j, \label{eq:FRW1}
\end{eqnarray}
%
%==========================%  
where $F^{-1}=(9/4)t_0^2(1+\omega )^2$ and $\omega=1/3,0$ correspond
to $\gamma=4/3,1$. In the coordinate system 
$(\Omega, x^i)$ the metric is singular at the timelike infinity 
$\Omega \to 0$. However, 
%========<Equation>========%
%
\begin{eqnarray}
n^a=\Omega^{-3\omega-5}{\hat g}^{ab}{\hat \nabla}_b \Omega 
=-F(\partial_\Omega)^a   
\label{eq:TN}
\end{eqnarray}
%
%==========================%  
and 
%========<Equation>========%
%
\begin{eqnarray}
q_{ab}=\Omega^2 a^2\gamma_{ij}dx^i dx^j=\gamma_{ij}dx^idx^j. \label{eq:FRW2}
\end{eqnarray}
%
%==========================%  
are smooth tensors at the infinity and contain all informations  
of gravitational field.  
From eq. (\ref{eq:TN}) the function $F$ can be 
expressed by $F=-\mbox{\pounds}_n \Omega $ which will be turned 
out to contain the information of the energy in Sec. V. 

We can see easily that the case of 
spatially open FRW space-times are accidentally combined to  
eq.(\ref{eq:FRW1}) $\sim$ (\ref{eq:FRW2}) with $\omega=-1/3$.

\section{AFRWTI space-times}

From the study of the completion of FRW space-times in 
the previous section, we may define space-times which are asymptotic
to radiation dominant FRW space-times at timelike infinity as follows;

{\it Definition}:A space-time (${\hat M}, {\hat g}_{ab}$) will be 
said to be {\it asymptotic to radiation dominant
spatially flat (or open) FRW space-times at timelike infinity} ${\cal
I}^+$(AFRWTI space-times) if there exists a smooth function 
$\Omega$ satisfying the following
features (i)(ii) and the energy momentum tensor satisfies the 
fall off condition (iii);

(i)$\Omega|_{{\cal I}^+}=0$ and $ d \Omega |_{{\cal I}^+} \neq 0$

(ii)The following quantities have smooth limits on ${\cal I}^+$. 
%========<Equation>========%
%
\begin{eqnarray}
n^a=\Omega^{-3\omega -5}{\hat g}^{ab}{\hat \nabla}_b \Omega
\end{eqnarray}
%
%==========================%  
%========<Equation>========%
%
\begin{eqnarray}
q_{ab}=\Omega^2({\hat g}_{ab}
+F^{-1}\Omega^{-3 \omega -5}{\hat \nabla}_a \Omega 
{\hat \nabla}_b \Omega)=\Omega^2{\hat q}_{ab}, 
\end{eqnarray}
%
%==========================%  
where $F=-\mbox{\pounds}_n \Omega$ and $\omega =1/3 $ for flat(or $\omega= -1/3$ 
for open). 

(iii)${\hat T}_{{\hat \mu}{\hat \nu}}:=
\h T_{ab}\base{\h e}{\mu}^a\base{\h e}{\nu}^b
= O( \Omega^4 )$ near ${\cal I}^+$,
where $\lbrace \base{\h e}{\mu}^a \rbrace_{\ssr \mu =0,1,2,3}$ is a tetrad of
the metric $\h g_{ab}$.

In the above definition, we excluded the matter dominant case ($\omega =0$) 
because we realise that one cannot obtain the comprehensive asymptotic 
structure. The detail will be discussed in below and Sec. VI. In
the above formulation, the $(3+1)$-decomposition split is implicitly
included. Since we are interested in asymptotic structure at timelike 
infinity, the treatment is plausible.

In terms of $(n^a, q_{ab})$ the extrinsic curvature of the spacelike 
hypersurface $\Omega=$constant is written as   
%========<Equation>========%
%
\begin{eqnarray}
{\hat K}_{ab}=\frac{1}{2}\mbox{\pounds}_{\hat n}{\hat q}_{ab}
=F^{1/2}\Omega^{\frac{1}{2}(3\omega -1)} q_{ab}
+\frac{1}{2}F^{-1/2}\Omega^{\frac{1}{2}(3\omega+1)}
\mbox{\pounds}_n q_{ab}
\end{eqnarray}
%
%==========================%  
Since the tensor is singular at $\Omega =0$, we define a 
smooth tensor $K_{ab}$ by  
%========<Equation>========%
%
\begin{eqnarray}
K_{ab}=\Omega^{-\frac{1}{2}(3\omega-1)}{\hat K}_{ab}
=F^{1/2}q_{ab}+\frac{1}{2}F^{-1/2}\Omega \mbox{\pounds}_n q_{ab}
\label{eq:K}
\end{eqnarray}
%
%==========================%  

The time-space components of the Einstein equation is written as 
%========<Equation>========%
%
\begin{eqnarray}
\Omega^{-\frac{1}{2}(3\omega +5)} \base{\h e}{I}^a{\hat n}^b {\hat G}_{ab}
=(D_aK^a_b-D_bK) \base{e}{I}^b, \label{eq:G0I} 
\end{eqnarray}
%
%==========================%  
where $({\hat n}^a, \lbrace \base{\h e}{I}^a \rbrace_{\ssr I =1,2,3})$ 
is a tetrad of the metric ${\hat g}_{ab}$ and $\lbrace \base{e}{I}^a 
\rbrace_{\ssr I=1,2,3}$ is a smooth triad of the metric $q_{ab}$. 
Substituting eq. (\ref{eq:K}) into (\ref{eq:G0I}) and imposing 
the fall-off condition (iii), 
we find $F{\hat =}{\rm constant}$, where `${\hat =}$' 
denotes the evaluation 
on ${\cal I}^+$.  By virtue of the remaining freedom of the 
conformal rescaling we can always set 
$F{\hat =}1$ without loss of generality. 

The space-space components of
the Einstein equation is written as 
%========<Equation>========%
%
\begin{eqnarray}
\Omega^{-2}{\hat R}_{ab}\base{\h e}{I}^a \base{\h e}{J}^b
&=& \Bigl[ {}^{3}R_{ab}-F^{1/2}D_aD_bF^{-1/2}
- \frac{1}{2}(3\omega-1)\Omega^{3 \omega +1}
F^{1/2}K_{ab} \nonumber \\
& & ~~~+\Omega^{3\omega +1}(-2K^c_bK_{ac}+K_{ab}
K)  \Bigr]\base{e}{I}^a \base{e}{J}^b.  \label{eq:R0I}
\end{eqnarray} 
%
%==========================%  
Substituting eq. (\ref{eq:K}) and $F{\hat =}1$ into 
eq.(\ref{eq:R0I}) 
and imposing the fall-off condition (iii), we find  
%========<Equation>========%
%
\begin{equation}
    {}^{3}R_{ab}{\hat =} 2 \k \gamma_{ab}
             \mmbox{3ex}{$\Leftrightarrow$}{3ex}
    \up{(0)}q_{ab} {\hat =} \gamma_{ab} \label{eq:Lemma}
\end{equation}
%
%==========================%
where $\k=0$ and $\k=-1$ for $\w =1/3~~{\rm and}~~\omega=-1/3$,
respectively. In other words, $q_{ab}$ is locally the metric 
of 3-dimensional Euclid and unit hyperboloid space at timelike infinity for 
the asymptotically flat and open FRW space-times, respectively. 
These features, $F\h=1$ and $\up{(0)}q_{ab} {\hat =} \gamma_{ab}$, 
belong to the 0th order structure in the words of the ref. \cite{Gen}.

In terms of $(n^a,q_{ab})$, the electric part of the Weyl
tensor is written as  
%========<Equation>========%
%
\begin{eqnarray}
{\hat E}_{ab}& =& F^{1/2}D_aD_bF^{-1/2}+\Omega^{3\omega +1}K_{ac}K^c_b
+\frac{1}{2}(3\omega -1)F^{1/2}\Omega^{3 \omega +1}K_{ab}
\nonumber \\
& & ~~~-F^{-1/2}\Omega^{3 \omega  +2}\mbox{\pounds}_n K_{ab}
+\frac{1}{2}({\hat q}^c_a{\hat q}^d_b-{\hat q}_{ab}{\hat n}^c{\hat n}^d)
{\hat L}_{cd},
\end{eqnarray}
%
%==========================% 
where ${\hat L}_{ab}={\hat R}_{ab}-\frac{1}{6}{\hat g}_{ab}
{\hat R}$. By using expansion series
$F=1+\up{(1)}F\Omega + \cdots$ and 
$\h L_{ab} \base{\h e}{\mu}^a \base{\h e}{\nu}^b=
\up{(3)}\h L_{\ssr {\h\mu\h\nu}}\, \Omega^3+
\up{(4)}\h L_{\ssr {\h\mu\h\nu}}\, \Omega^4+\cdots$, 
which follows from the condition (iii), around $\Omega=0$,
we obtain the expression 
%========<Equation>========%
%
\begin{eqnarray}
{\hat E}_{ab}=-\frac{1}{2}\Omega(D_aD_b+\k q_{ab})\up{(1)}F
+ \frac12 \base{e}{I}_{a}\base{e}{J}_b 
\Omega( \up{(3)}\h L_{\ssr{\h I\h J}}
                     +\delta_{\ssr{I,J}}\!\!\up{(3)}\h
L_{\ssr{\h0\h0}}) 
+O(\Omega^2).
\end{eqnarray}
%
%==========================% 
Although the condition (iii) implies $\up{(3)}\h
L_{\ssr{\h\mu\h\nu}}=0$ we leave the term $\up{(3)}\h L_{\ssr
{\h\mu\h\nu}}$ in the above equation in order to give 
a comment on the matter dominant universes below. 
To discuss the leading behaviour of gravitational field we 
define the tensor field, $E_{ab}:={\hat E}_{ab}\Omega^{-1}$. 
This tensor field $E_{ab}$ satisfies the field equation 
%========<Equation>========%
%
\begin{equation}
     D_{[a}E_{b]c}{\hat =} \frac12 D_{[a}\left[
                \base{e}{I}_{b]}\base{e}{J}_c ( \up{(3)}\h L_{\ssr{\h I\h J}}
                     +\delta_{\ssr{I,J}}\!\!\up{(3)}\h L_{\ssr{\h0\h0}})
                                        \right].\label{eq:EL}
\end{equation}
%
%==========================%
From the tracelessness of $E_{ab}$ we find that $\up{(1)}F$ obeys 
the differential equation,
%========<Equation>========%
%
\begin{eqnarray}
  (D^2+3\k)\up{(1)}F{\hat =} 2\up{(3)}\h L_{\ssr{\h0\h0}} 
                        +\frac12 \up{(3)}\h L_{\ssr{\h I\h I}}.
                 \label{eq:F=S}
\end{eqnarray}
% 
%==========================% 
In AFRWTI space-times,
$\up{(3)}\h L_{\ssr {\h\mu\h\nu}}$ vanishes on ${\cal I}^+$, and  
we obtain 
%========<Equation>========%
%
\beqn
     & E_{ab}\h=-\frac{1}{2}(D_aD_b+\k q_{ab})\up{(1)}F \label{eq:EF} & \\
     &  D_{[a}E_{b]c}{\hat =} 0 
          \mmbox{3ex}{$\Rightarrow$}{3ex} 
        D_a E^{ab}\h=0  \label{eq:Bian} & \\
     &  (D^2+3\k)\up{(1)}F{\hat =}0 & \label{eq:F}
\end{eqnarray}
%
%==========================%
The behaviours of the function $\up{(1)}F$ and the tensor
field $E_{ab}$ are completely determined regardless of 
the leading behaviour of the
energy-momentum tensor $\up{(4)}\h L_{\ssr {\h\mu\h\nu}}$. 
These features belong to the 1st order structure in the 
words of the ref. \cite{Gen}. 

If one wants to consider asymptotically matter dominant FRW 
space-times($\w =0$), one should demand ${\hat T}_{{\hat \mu}
{\hat \nu}}=O(\Omega^3)$ instead of the fall-off condition (iii) 
because of the behaviour of ${\hat T}_{{\hat \mu}{\hat \nu}}$ in the 
exact matter dominant FRW space-time. In this case we obtain 
$F{\hat =}1$ and eq. (\ref{eq:Lemma}) but not eqs.(\ref{eq:EF})
--(\ref{eq:F}). That is, asymptotically matter dominant FRW
space-times  
possess 0th order asymptotic structure but not the 1st order
structure.  $\up{(1)}F$ depends on the behaviour of $\up{(3)}\h L_{\ssr
{\h\mu\h\nu}}$. 
Among them the case with $\up{(3)}\h T_{\ssr {\h\mu\h\nu}} =\rho_0 \delta_{\h 0
\h\mu}\delta_{\h 0 \h \nu}$, where $\rho_0={\rm const.}$ may be worth
discussing here. In this case 
%========<Equation>========%
%
\beqn
     & E_{ab}\h=-\frac{1}{2}(D_aD_b+\k
     q_{ab})\up{(1)}F+\frac{1}{6}\rho_0
      q_{ab} \label{eq:Mufu1} & \\
     &  D_{[a}E_{b]c}{\hat =} 0 
          \mmbox{3ex}{$\Rightarrow$}{3ex} 
        D_a E^{ab}\h=0  & \\
     &  (D^2+3\k)\up{(1)}F{\hat =}\rho_0 & \label{eq:Mufu2}
\end{eqnarray}
%
%==========================%
hold instead of eqs. (\ref{eq:EF})--(\ref{eq:F}). 

\section{Local Asymptotic Symmetry}

The FRW space-times have a timelike conformal Killing vector (CKV) 
field $\h \xi_{(t)}=\partial_\eta$ and 
spacelike Killing vector fields. Here, we seek its asymptotic 
correspondences in AFRWTI space-times. Since CKV 
$\h \xi$ is defined by $\mbox{\pounds}_{\h \xi}\h g_{ab}=f\h g_{ab}$, 
where $f$ is a smooth function, CKV is a Killing
vector field of a conformal metric $g_{ab}=\Omega^2 \h g_{ab}$ 
with $f =-2\mbox{\pounds}_{\h \xi}{\rm ln}\Omega $, that is, 
$\mbox{\pounds}_{\h \xi}g_{ab}=0$.

Simple calculations lead us to 
%========<Equation>========%
%
\begin{eqnarray}
& & \base{e}{0}^a \base{e}{0}^b
\mbox{\pounds}_{\hat \xi}(\Omega^2{\hat g}_{ab})
=3(1+\omega)\Omega^{-1} \mbox{\pounds}_{\hat \xi}\Omega -F^{-1}
\mbox{\pounds}_{\hat \xi}F+2F^{-1}\mbox{\pounds}_{[n,{\hat \xi}]}\Omega  \\
& & \base{e}{I}^a \base{e}{0}^b
\mbox{\pounds}_{\hat \xi}(\Omega^2{\hat g}_{ab})
=\Bigl[ F^{-1/2}D_a(\mbox{\pounds}_{\hat \xi}\Omega)
-\Omega^{\frac{3}{2}(1+\omega)}F^{-1/2}[{\hat \xi}, n]_a \Bigr]
\base{e}{I}^a  \\
& & \base{e}{I}^a \base{e}{J}^b 
\mbox{\pounds}_{\hat \xi}(\Omega^2{\hat g}_{ab})
= \base{e}{I}^a \base{e}{J}^b
\mbox{\pounds}_{\hat \xi}q_{ab},
\end{eqnarray}
%
%==========================% 
where ${\hat \xi}$ is a vector field and 
$\lbrace \base{e}{\mu}^a \rbrace_{\ssr \mu =0,1,2,3}$ 
is a tetrad of the metric $g_{ab}$. 

The vector field which induces the timelike translation in 
the conformally transformed space-times with metric
$g_{ab}=\Omega^2{\hat g}_{ab}$ 
can be written as $ {\hat \xi}=\alpha \Omega^{\frac{3}{2}(1+\omega)} 
\partial_\Omega $, where $\alpha$ is a smooth function.  
Using this expression we see that the above equations become
%========<Equation>========%
%
\begin{eqnarray}
& & \base{e}{0}^a \base{e}{0}^b 
\mbox{\pounds}_{\hat \xi}(\Omega^2{\hat g}_{ab})
= \Omega^{\frac{3}{2}(\omega +1)}[\up{(0)}\alpha \up{(1)}F-2
\up{(1)}\alpha ] +\cdots\\
& & \base{e}{I}^a \base{e}{0}^b
\mbox{\pounds}_{\hat \xi}(\Omega^2{\hat g}_{ab}) 
=\Omega^{\frac{3}{2}(\omega +1)}D_{\ssr I} \up{(0)}\alpha +\cdots\\
& & \base{e}{I}^a \base{e}{J}^b 
\mbox{\pounds}_{\hat \xi}(\Omega^2{\hat g}_{ab})
= \up{(0)}\alpha \Omega^{\frac{3}{2}(\omega +1)} \up{(1)}q_{\ssr IJ} +\cdots,
\end{eqnarray}
%
%==========================% 
where we used the expansion series $\alpha=\up{(0)}\alpha+\Omega 
\up{(1)}\alpha+\cdots$ and $q_{ab}=\gamma_{ab}+\Omega
\up{(1)}q_{ab}+\cdots$ around $\Omega=0$. 
Hence, $\base{e}{\mu}^a \base{e}{\nu}^b 
\mbox{\pounds}_{\hat \xi}(\Omega^2{\hat g}_{ab})
=O(\Omega^{\frac{3}{2}(\omega+1)})$ holds in general. Thus, 
the vector field defined by ${\hat \xi}=\alpha \Omega^{\frac{3}{2}(1+\omega)} 
\partial_\Omega $ deserves to be called asymptotically timelike 
conformal Killing vector fields. 

On the other hand, the vector field which induces the spacelike translation in the 
physical space-times with the metric ${\hat g}_{ab}$ can be 
written as ${\hat \xi}^a=\xi^{\ssr I} \base{e}{I}^a$. In this case the Lie 
derivatives of the physical metric become
%========<Equation>========%
%
\begin{eqnarray}
& & {\hat n}^a {\hat n}^b\mbox{\pounds}_{\hat \xi} {\hat g}_{ab}
= -\Omega \mbox{\pounds}_{\hat \xi}\up{(1)}F +\cdots\\
& & \base{\h e}{I}^a {\hat n}^b\mbox{\pounds}_{\hat \xi} {\hat g}_{ab} 
=-\Omega^{\frac{3}{2}(\omega +1)}[{\hat \xi},n]_{\ssr I} +\cdots\\
& & \base{\h e}{I}^a \base{\h e}{J}^b \mbox{\pounds}_{\hat \xi} {\hat g}_{ab}
= \base{e}{I}^a \base{e}{J}^b
\mbox{\pounds}_{\hat \xi}\up{(0)}q_{ab}+\cdots.
\end{eqnarray}
%
%==========================%
The vector field which generates the isometry always exists in 
the hypersurface with the metric $\gamma_{ab}$. If we adopt 
${\hat \xi}$ as such vector fields,  
$ \base{\h e}{\mu}^a \base{\h e}{\nu}^b 
\mbox{\pounds}_{\hat \xi}({\hat g}_{ab})
=O(\Omega)$ holds.  Thus, AFRWTI space-times have asymptotic 
spacelike Killing vector fields. 

We can obtain more comprehensive results for cases with 
$\omega=-1/3$ because such cases is belonging to 
space-times defined in ref.\cite{Gen}. 
See the ref. \cite{Gen} for the detail.

\section{Energy}

In this section we define the total 
energy of AFRWTI space-times from the electric part of the Weyl 
tensor. The coefficient 
of the monopole component of $\up{(1)}F$, $a_{00}$, is expected to 
express the total conserved `energy' because 
$E_{ab}$, which plays the tidal force in the equation of geodesic
congruence, is written in terms of $\up{(1)}F$ as eq. (\ref{eq:EF}). When one 
considers the total energy, one need not take account of 
the magnetic part of the Weyl tensor. The part contains the
information of the total angular momentum or/and the local energy of a sort of 
gravitational wave.

First of all, we consider $\omega=1/3$ case. In this case, 
the `metric' of the timelike infinity ${\cal I}^+$ becomes the 
Euclid metric 
%========<Equation>========%
%
\begin{eqnarray}
q_{ab}{\hat =}(dr)_a(dr)_b+r^2 [(d\theta)_a (d\theta)_b
+{\rm sin}^2 \theta (d\phi)_a (d\phi)_b].
\end{eqnarray}
%
%==========================%
From the regularity condition on the spatial infinity
($r \rightarrow \infty$), the general solution of
eq. (\ref{eq:F}) with $\k=0$ is given by 
%========<Equation>========%
%
\begin{eqnarray}
\up{(1)}F=\sum_{\ell, m} r^{-(\ell +1)}a_{\ell m}Y_{\ell m }, \label{eq:FFF1}
\end{eqnarray}
%
%==========================%
Substituting eq. (\ref{eq:FFF1}) into eq. (\ref{eq:EF}), we obtain 
%========<Equation>========%
%
\begin{eqnarray}
\int dSE_{rr}=\frac{16\pi}{r}a_{00}. 
\end{eqnarray}
%
%==========================%
Since the `energy' is expected to be contained in the monopole component 
of $E_{ab}$, we may define the energy, ${\cal E}$, as 
%========<Equation>========%
%
\begin{eqnarray}
{\cal E}=a_{00}=\frac{r}{16\pi}\int dSE_{rr}
\end{eqnarray}
%
%==========================%

Second, let us consider $\omega =-1/3$ case. In this case, 
the `metric' of the timelike infinity ${\cal I}^+$ becomes 
%========<Equation>========%
%
\begin{eqnarray}
q_{ab}{\hat =}(d\chi)_a(d\chi)_b+{\sinh}^2\chi [(d\theta)_a (d\theta)_b
+{\rm sin}^2 \theta (d\phi)_a (d\phi)_b ]
\end{eqnarray}
%
%==========================%
From the regularity condition on the spatial infinity
($\chi \rightarrow \infty$), the general solution of
eq. (\ref{eq:F}) with $\k=-1$ is given by 
%========<Equation>========%
%
\begin{eqnarray}
\up{(1)}F=\sum_{\ell ,m} a_{\ell m}\frac{1}{{\sqrt {{\rm sinh}\chi}}}
P^{\ell+\frac{1}{2}}_{\frac{3}{2}}({\rm cosh}\chi)Y_{\ell m},
\end{eqnarray}
%
%==========================%
where $P^{\ell+\frac{1}{2}}_{\frac{3}{2}}({\rm cosh}\chi)$ denotes the 
associated Legendre 
function\cite{Gen}. In the same way as $\omega=1/3$, the `energy' may
be defined as 
%========<Equation>========%
%
\begin{eqnarray}
{\cal E}=a_{00}=\frac{{\rm sinh}\chi}{8\pi}\int E_{\chi \chi}dS.
\end{eqnarray}
%
%==========================%
Note that 
this expression is same as the energy defined in ref. \cite{Gen}.  
If $ \up{(3)}\h T_{\ssr {\h\mu\h\nu}} =\rho_0 \delta_{\h 0
\h\mu}\delta_{\h 0 \h \nu}$ with $\rho_0=$const., using eqs. 
(\ref{eq:Mufu1})-(\ref{eq:Mufu2}) we see 
$\up{(1)}F=\sum_{\ell, m} r^{-(\ell +1)}a_{\ell m}Y_{\ell m }+
\frac{\rho_0}{6r^2}$ for $\omega=1/3$ and 
$\up{(1)}F=\sum_{\ell ,m} a_{\ell m}\frac{1}{{\sqrt {{\rm sinh}\chi}}}
P^{\ell+\frac{1}{2}}_{\frac{3}{2}}({\rm cosh}\chi)Y_{\ell m}
-\frac{1}{3}\rho_0$ for $\omega=-1/3$, which lead us to the result 
${\cal E}=a_{00}$.

\section{Summary and Discussion}

In this paper, we defined space-times which are asymptotic to 
radiation dominant universes and showed that time slices of 
these space-times approach conformally to Euclid space or hyperboloid 
space. Furthermore, we found that the asymptotic 
timelike conformal Killing vector and spacelike Killing vector fields 
exist locally. 
Finally, we defined the energy from the electric part of the Weyl tensor. 
Although we have obtained successfully the asymptotic structure, we should 
mention that the strong assumption on the energy-momentum tensor, 
that is, ${\hat T}_{\h \mu \h \nu}=O(\Omega^4)$, was needed. 

In our definition, we needed to exclude matter dominant universe
because the energy-momentum tensor may not be expect to 
decay faster than ${\hat T}_{{\hat \mu}
{\hat \nu}}
=O(\Omega^3)$. In this case, as we said in Sec. III, 
the energy-momentum tensor term 
leaves in the expression of $E_{ab}$ and we must solve Einstein
equation to obtain the detail of the asymptotic structure. 
This is hard task. Moreover, one needs to  
carefully consider even radiation dominant universe. From the
educational point of view, we consider the linear perturbation on 
FRW space-time. In this case, $E_{ab}$ contains the term $ 
{\hat T}_{\ssr {\hat 0 \hat 0}} \sim \rho (1+\delta)$, 
where $\delta =\delta \rho/\rho$. As $\delta $ satisfies 
the equation ${\dot \delta} \simeq ({\dot a}/a)\delta$ and thus  
$\delta \propto a \propto \Omega^{-1}$, the leading term has 
the order of $\Omega^3$. Hence 
${\cal I}^+$ defined by us can not contain the region, where the perturbation 
exists. We remind you, however, our 
approach may be extended to more 
generalised Robertson-Walker space-times using 
inhomogeneous fibre\cite{GRW}. 

Finally, we give a comment on a relation between 
gravitational instanton and the energy defined in the previous 
section. It is likely that the instanton corresponds to  
the `ground state'. Hawking and Horowitz discussed that the physical 
Hamiltonian 
$H_P$ should be finite and given by $H_P=H-H_0$, 
where $H_0$ is the Hamiltonian of the background space-time that  
is stationary or static\cite{HH}. From the definition we naively 
see that 
the ground state with the minimum energy is the background
space-time. Given a solution, only the surface term 
in $H_P$ lefts and then $H_P=-(1/8\pi)\int_{S_\infty}dS(k-k_0)$, 
where $k$ and $k_0$ are the trace of the extrinsic curvatures 
of $S_\infty$ in full and background space-times; $S_\infty$ 
is a 2-dimensional surface at infinity. Here we 
adopted the gauge such that the lapse function $N=1$ and the shift 
vector $N^a=0$. The value supplies the `total energy'. 
In particular, this is the same as ADM energy 
for asymptotically flat and Abbott-Deser energy\cite{AD} for 
asymptotically anti-deSitter space-times. The each ground state 
is Minkowski and anti-deSitter space-times. 
On the other hand, it is likely that our energy ${\cal E}$ 
defined in the previous section coincides to the case 
in which the background space-times is FRW space-times. In fact, 
${\cal E}$ has approximately the expression ${\cal E} \sim 
\int \delta \rho {\sqrt {\h q}}d^3x$ for the linear perturbation 
in the flat FRW universe. The plausible construction 
of the physical Hamiltonian of the dynamical background space-times 
should be established in order to study deeply non-linear version.

\section*{Acknowledgements}
We would like to thank K. Sato, M. Sasaki and Y. Suto for their 
encouragements. TS thanks G.W. Gibbons for valuable discussion. 
We are grateful to anonymous referees for their useful comments. 
We also thanks M. Spicci for his careful reading of this manuscript. 
This work is partially supported by JSPS[No.310(T.S.)]. 

\appendix

\section{}

We give two examples of AFRWTI space-times which belong to
Bianchi type I and V. The matter components is perfect fluid having the 
equation of state, $P=(1/3)\rho$. 

First, we take an 
exact solution of Bianchi type I found in
\cite{BianchiI}\cite{General}. The metric is given by 
%========<Equation>========%
%
\begin{eqnarray}
ds^2=-[A(t)]^{2/3}dt^2+t^{2p_1}[A(t)]^{4/3-2p_1}dx^2+
t^{2p_2}[A(t)]^{4/3-2p_2}dy^2+t^{2p_3}[A(t)]^{4/3-2p_3}dz^2
\end{eqnarray}
%
%==========================%
where $A(t)=(\alpha+\beta t^{2/3})^{3/2}$ 
and $p_1,p_2,p_3$ are constants satisfying $p_1+p_2+p_3=1$ and 
$p_1^2+p_2^2+p_3^2=1$. The energy density is 
%========<Equation>========%
%
\begin{eqnarray}
\rho=\frac{4\beta}{3t^{4/3}A^{4/3}}.
\end{eqnarray}
%
%==========================%

We define the function $\Omega$ as follows,
%========<Equation>========%
%
\begin{eqnarray}
\Omega:=t^{-2/3}
\end{eqnarray}
%
%==========================%
In terms of the new coordinate $\Omega$, the metric is written as 
%========<Equation>========%
%
\begin{eqnarray}
ds^2=-F^{-1}\Omega^{-6}(d\Omega)^2+\Omega^{-2}
\Bigl[ (\beta+\alpha \Omega)^{2-3p_1}dx^2+(\beta+\alpha \Omega)^{2-3p_2}dy^2
+(\beta+\alpha \Omega)^{2-3p_3}dz^2 \Bigl],
\end{eqnarray}
%
%==========================%
where
%========<Equation>========%
%
\begin{eqnarray}
F=\frac{4}{9}\frac{1}{\beta+\alpha \Omega}.
\end{eqnarray}
%
%==========================%
The energy density is written as 
%========<Equation>========%
%
\begin{eqnarray}
\rho=\frac{4\beta}{3(\beta+\alpha \Omega)^2}\Omega^4.
\end{eqnarray}
%
%==========================%
It is easy to see that the above solution satisfies the condition 
(i) $\sim$ (iii) of AFRWTI space-times with $\omega=1/3$. 

Next we consider an 
exact solution of Bianchi type V found in
\cite{BianchiV}\cite{General}. The metric
is given by 
%========<Equation>========%
%
\begin{eqnarray}
ds^2=A(t)B(t)\Bigl[ -dt^2+dx^2+e^{2x}\Bigl\lbrace 
\Bigl(\frac{A}{B} \Bigr)^{{\sqrt {3}}}dy^2
+\Bigl( \frac{B}{A}\Bigr)^{{\sqrt {3}}}dz^2 \Bigr\rbrace \Bigr],
\end{eqnarray}
%
%==========================%
where $A={\rm sinh}t$ and $B=\alpha {\rm cosh}t+\beta{\rm sinh}t$ and 
the energy density is 
%========<Equation>========%
%
\begin{eqnarray}
\rho=\frac{3\beta}{(AB)^2}
\end{eqnarray}
%
%==========================%

We define the function $\Omega$ as follows, 

%========<Equation>========%
%
\begin{eqnarray}
\Omega:=e^{-t}.
\end{eqnarray}
%
%==========================%
In term of $\Omega$, the metric is written as 
%========<Equation>========%
%
\begin{eqnarray}
ds^2 & = & -F^{-1}\Omega^{-4}(d\Omega)^2 \nonumber \\
& & ~~~~+\Omega^{-2}
\Bigl[  \frac{1}{4}(1-\Omega^2)(\alpha_++\alpha_-\Omega^2)dx^2
+\frac{1}{4}e^{2x}\frac{(1-\Omega^2)^{{\sqrt {3}}+1}}{(\alpha_++\alpha_-\Omega^2)
^{{\sqrt {3}}-1}} \Bigl\lbrace  dy^2 
+\Bigl( \frac{\alpha_++\alpha_-\Omega^2}{1-\Omega^2} \Bigr)^3 dz^2 
\Bigr\rbrace \Bigr],
\end{eqnarray}
%
%==========================%
where $\alpha_\pm=\alpha \pm \beta$ and 
%========<Equation>========%
%
\begin{eqnarray}
F=\frac{4}{(1-\Omega^2)(\alpha_++\alpha_-\Omega^2)}.
\end{eqnarray}
%
%==========================%
The energy density is 
%========<Equation>========%
%
\begin{eqnarray}
\rho=\frac{48\beta}{(1-\Omega^2)^2(\alpha_++\alpha_-\Omega^2)^2}\Omega^4.
\end{eqnarray}
%
%==========================%
We see again that the solution satisfies the condition 
(i) $\sim$ (iii) of AFRWTI space-times with $\omega=-1/3$.

\end{document}